\documentclass[doublecol]{epl2}
\usepackage{amsmath}
\usepackage{amssymb}
\usepackage{graphicx}
\usepackage{dcolumn}
\usepackage{bm}
\usepackage{subfigure}
\usepackage{changes}

\title{Disordered configurations of the Glauber model in \\ two-dimensional networks}
\shorttitle{Disordered configurations of the Glauber model in two-dimensional networks}

\author{Iva Ba\v{c}i\'c \inst{1} \and Igor Franovi\'c\inst{1} \and Matja{\v z} Perc\inst{2,3,4}}
\shortauthor{Iva Ba\v{c}i\'c, Igor Franovi\'c and Matja{\v z} Perc}
\institute{\inst{1} Scientific Computing Laboratory, Center for the Study of Complex Systems,
Institute of Physics Belgrade, University of Belgrade, Pregrevica 118, 11080 Belgrade, Serbia\\
  \inst{2} Faculty of Natural Sciences and Mathematics, University of Maribor, Koro{\v s}ka cesta
160, SI-2000 Maribor, Slovenia\\
  \inst{3} CAMTP -- Center for Applied Mathematics and Theoretical Physics, University
of Maribor, Mladinska 3, SI-2000 Maribor, Slovenia\\
  \inst{4} Complexity Science Hub, Josefst{\"a}dterstra{\ss}e 39, A-1080 Vienna, Austria}


\abstract{
We analyze the ordering efficiency and the structure of disordered configurations for the zero-temperature Glauber model on Watts-Strogatz networks obtained by rewiring 2D regular square lattices. In the small-world regime, the dynamics fails to reach the ordered state in the thermodynamic limit. Due to interplay of the perturbed regular topology and the energy neutral stochastic state transitions, the stationary state consists of two intertwined domains, manifested as multi-clustered states on the original lattice. Moreover, for intermediate rewiring probabilities, one finds an additional source of disorder due to the low connectivity degree, which gives rise to small isolated droplets of spins. We also examine the ordering process in paradigmatic two-layer networks with heterogeneous rewiring probabilities. Comparing the cases of a multiplex network and the corresponding network with random inter-layer connectivity, we demonstrate that the character of the final state qualitatively depends on the type of inter-layer connections.
}

\begin{document}

\maketitle

The interplay of local dynamics and the underlying network topology has been in the focus of research in physics and various interdisciplinary fields \cite{albertbarabasi2002,strogatz2001,boccaletti2006}, having recently attracted considerable interest in the context of phase ordering processes \cite{bray2002,castellano2009,pastorsatorras2015}.
The Ising-Glauber model \cite{krapivsky2010} constitutes one of the paradigmatic models for analyzing such processes \cite{castellano2005}. While it has been introduced to describe the nonequilibrium dynamical behavior of magnetic systems consisting of a large number of interacting particles, it has since been applied to a variety of other problems, including those in social sciences \cite{stauffer2008}, geology \cite{ganguly2012}, and electrochemistry \cite{bosco1993}.

Within the Glauber model, the spin variables can assume two discrete values, having the states of nodes evolve according to the local majority rule. The Glauber model was initially defined on a regular lattice \cite{krapivsky2010}. Nevertheless, given that non-lattice topologies including random, scale-free \cite{barabasialbert1999} and small-world \cite{wattsstrogatz1998} networks are often better suited to describe real-world systems, the issue of Glauber dynamics on complex networks has been gaining increasing attention \cite{castellano2005,castellano2006,herrero2009,baek2012}. Apart from such models, complexity of interactions in many real-world systems may also involve "networks of networks" featuring modular or multilayer architecture \cite{boccaletti2014}, the scenarios which have been much less explored in the framework of Glauber dynamics. 

Our work addresses two problems of ordering in complex networks: (i) the disordered states of the zero-temperature Glauber model on monolayer rewired networks, where we identify two types of disordered configurations, and (ii) the ordering process on two-layer rewired networks, where we find that the ordering process is strongly affected by the type of inter-layer connections.

\begin{figure}[t]
\centering
\includegraphics[width=8cm]{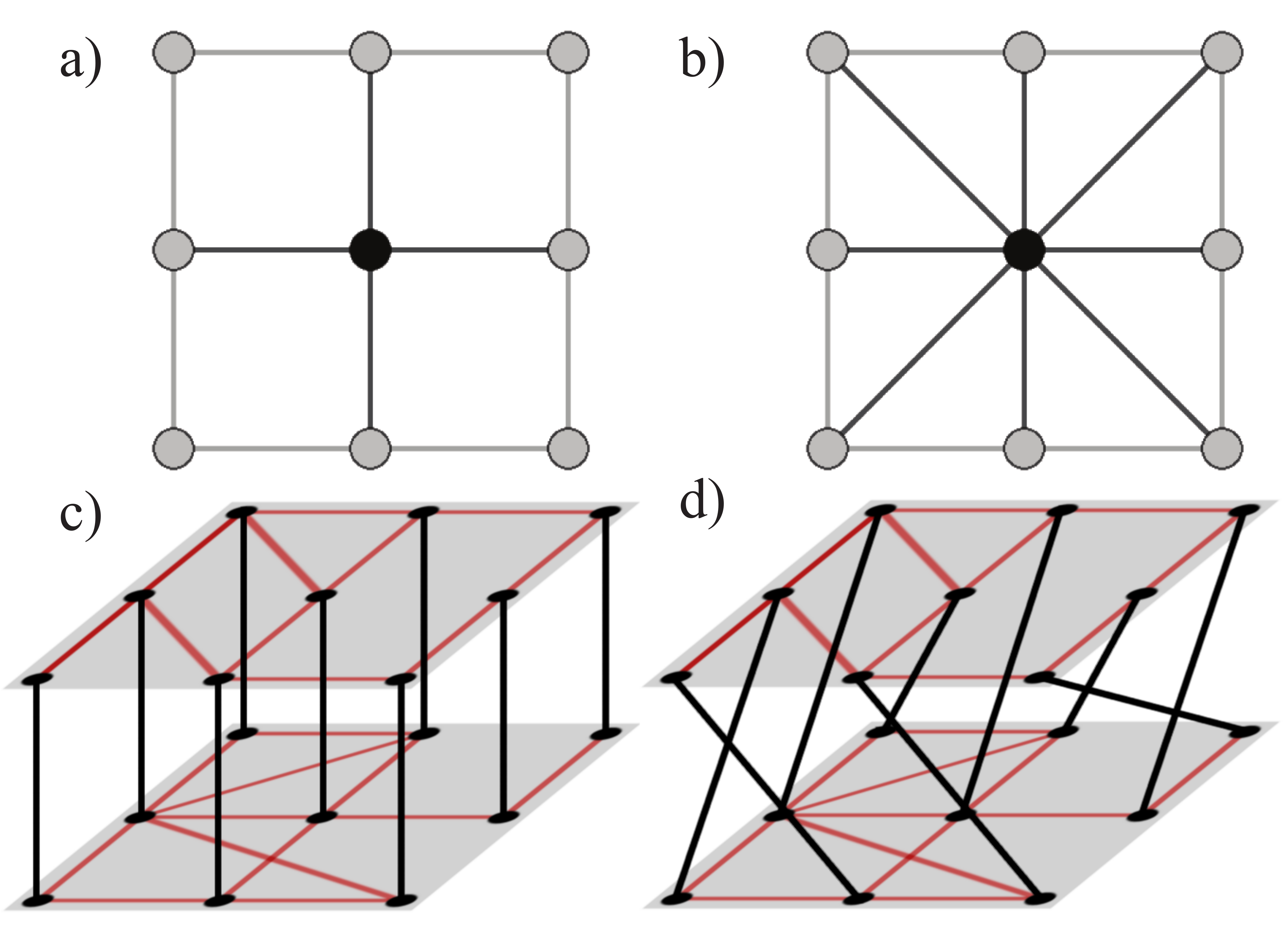}
\caption{(Color online) Considered network topologies. (a) shows the scenario of a monolayer network with  nearest-neighbor interactions $(k=4)$, whereas (b) illustrates the case where the next-nearest neighbor interactions are also included $(k=8)$; (c) concerns the case of a multiplex two-layer network, whereas (d) refers to scenario with random connectivity between the two layers.} \label{fig:lattice_neighbors}
\vspace{-0.5cm}
\end{figure}

In case of the two-dimensional square lattice, when only interactions between four nearest neighbors are taken into account, see Fig.  \ref{fig:lattice_neighbors}(a), the zero-temperature Glauber dynamics is multistable \cite{spirin2001}. In particular, the system either reaches the ground state for $\approx 2/3$ of all the process realizations, or ends up in the frozen striped state with probability $p_f\approx 1/3$. Concerning rewired square lattices with coordination number $\langle k \rangle =4$, it has been shown that the dynamics fails to reach the ground state \cite{boyer2003,herrero2009}, but little is known about the nature of the associated disordered configurations.

Our immediate goals are to understand why the Glauber model on small-world networks fails to reach ground state and to gain insight into the character of the disordered states on rewired networks with $\langle k \rangle=4$ and
$\langle k \rangle=8$. We also study the ordering process on two-layer rewired networks with
$\langle k \rangle=4$, comparing the effects of different types of inter-layer connectivity, including multiplexing, see Fig. \ref{fig:lattice_neighbors}(c), and the scenario with connections distributed between randomly selected pairs of nodes, cf. Fig. \ref{fig:lattice_neighbors}(d).

\begin{figure*}[t]
\centering
\includegraphics[scale=0.6]{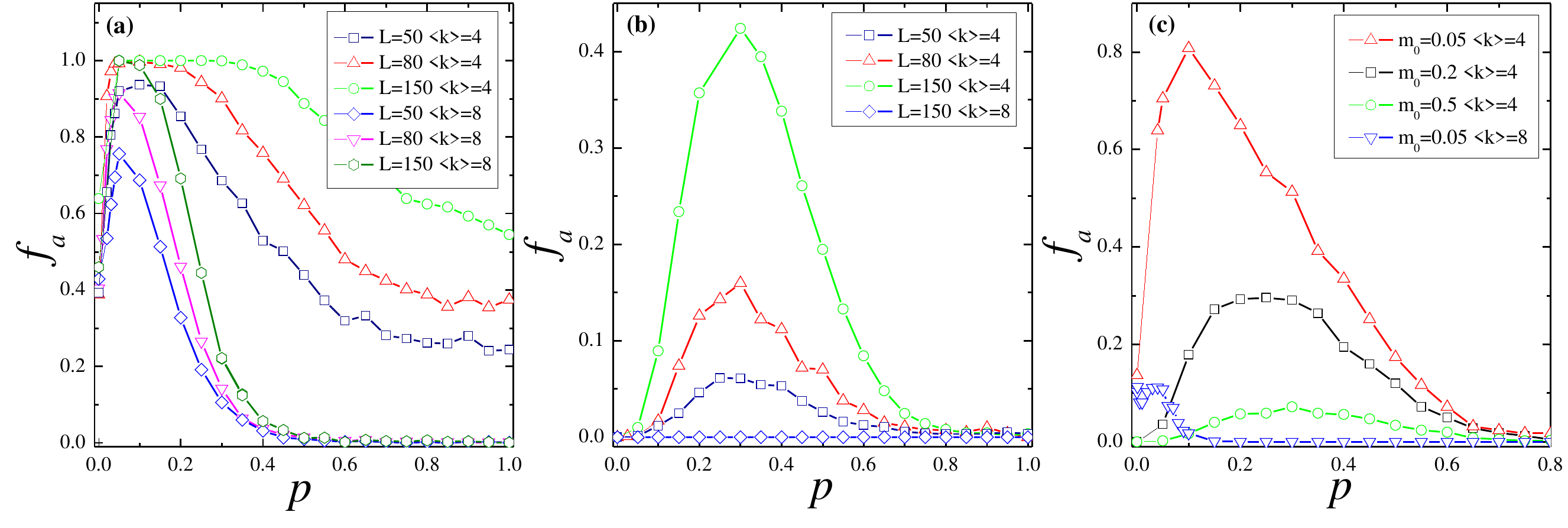}
\caption{(Color online) (a) Final fraction of active runs $f_a$ in terms of rewiring probability $p$ for the standard Glauber rule with $m_0=0$. The results are provided for networks with $\langle k \rangle=4$ and
$\langle k \rangle=8$ neighbors and $L\in\{50,80,150\}$. Note that complete ordering is not observed in the small-world regime $0<p \ll 1$ independent of $\langle k \rangle$. (b) Impact of modified Glauber rule: for
$\langle k \rangle=8$, the system reaches complete ordering (hence only the curve corresponding to $L=150$ is shown), whereas for $\langle k \rangle=4$, the frustration effect emerges at intermediate $p$, becoming more pronounced with the network size. (c) displays $f_a$ for systems governed by the standard Glauber rule starting from initial conditions $m_0 \neq 0$. The influence of small-worldliness is such that it suppresses disorder regardless of
$\langle k \rangle$ with increasing $m_0$, while it still promotes disorder at intermediate $p$ range for $\langle k \rangle=4$.} \label{fig:fa_vs_p}
\vspace{-0.4cm}
\end{figure*}

\textit{Model -- }In the Glauber model, the interactions are usually confined (but not necessarily restricted) to nearest-neighboring units. Incorporating higher-order competing (frustrated) interactions is one of the classical scenarios for the onset of new phases and potentially new types of phase transitions lying outside of Ising universality class. While the ferromagnetic interactions imposed by the model favor parallel alignment of spins, thermal noise prevents the system from reaching the ground state at any nonzero value of temperature. To avoid such stochastic effects which prevent full ordering, we consider systems quenched from an infinitely high temperature to absolute zero, in which spin states are initially uncorrelated and the net magnetization is vanishing. The Hamiltonian of the system is given by
$H = - \displaystyle\sum_{\langle ij \rangle} J_{ij}  S_i S_j$
where $S_i = \pm 1$  are Ising spin variables, the sum $\langle ij \rangle$ is over pairs of neighbors, and
$J_{ij} > 0$ are ferromagnetic coupling constants, assumed to be uniform in our paper ($J_{ij}=J$). Each pair of parallel neighboring spins contributes $-J$ to the energy, while the contribution of antiparallel pairs is $+J$. Without loss of generality, we set $J=1$ in the present study.

The state of the system evolves according to the majority rule applied to spins sequentially selected at random in each time step. This dynamical rule allows only energy lowering or the energy neutral state transitions. The former correspond to events where the spin variable is updated to the state prevalent in its local neighborhood, while the latter conform to scenario without a local majority, such that the given spin evolves stochastically with both orientations being equally likely.

Watts and Strogatz \cite{wattsstrogatz1998} have introduced an algorithm for generating small-world and random graphs by gradually rewiring a regular lattice. In their model, links from the regular lattice are chosen at random and replaced with new ones until a desired fraction of links $p$ is rewired. Rewiring effectively introduces shortcuts between distant nodes, thereby drastically reducing the mean shortest path even in the limit $p \rightarrow 0$. By increasing the amount of disorder ($p \rightarrow 1$) one obtains a random network with the mean connectivity conserved. Small-world networks are generated by introducing an intermediate level of disorder ($0<p \ll 1$),
and are characterized by the high clustering coefficient and the short average path length. The former implies that neighboring nodes tend to group in well connected clusters, whereas the latter means that an arbitrary distant node
can be reached by a small number of intermediate links.

We simulate Glauber dynamics of Ising spins on Watts-Strogatz rewired networks generated from two-dimensional regular $L \times L$ lattices with periodic boundary conditions. To understand the interplay between topological effects and the local majority dynamical rule, we vary several parameters in addition to $L$ and the rewiring probability $p$, including the mean connectivity degree $\langle k \rangle$ and the initial magnetization $m_0$. As an additional ingredient, we also examine how the ordering process is affected by whether the Glauber dynamical rule allows for stochastic flipping or not. We refer to the rule without stochastic flipping as the modified Glauber rule.

To distinguish the influence of rewiring itself from the effect of connectivity of the network, we compare the results of simulations on networks with $\langle k \rangle=4$ and $\langle k \rangle=8$ in the small-world regime. We regard the next nearest neighbors as first neighbors in the topological sense by setting all interactions to be of equal strength. Assigning a finite value to the initial magnetization $m_0 \neq 0$ can be understood as introducing
an initial bias toward local state clustering in the network. Modifying the Glauber dynamical rule by allowing state transitions only in the case of a strong local majority allows us to understand the effect of energy neutral processes on ordering in disordered topologies. In this scenario, nodes with an equal number of neighbors in both states are ignored when encountered during a trial rather than having their state determined stochastically. It turns out that the ground state is always reached on regular square lattices when a strong majority is necessary for state transition, i.e. the striped state turns out to be the consequence of energy neutral stochastic flips.

To gain a more comprehensive insight into structure of the disordered configurations, we make a distinction between the domains comprised of topologically connected nodes in the same state, and the clusters with respect to positions of the nodes on the original regular lattice. The lattice and the graph neighborhoods are always identical for spins placed on regular lattices. However, as the lattice structure is modified such that the links between neighbors are replaced by links to distant nodes, the lattice and the topological neighbors may not necessarily coincide, which results in rich patterns on the lattice. In order to investigate the crossover from frozen striped configurations occurring in regular lattices to disordered states occurring in the rewired lattices, we compare the correlation length $\xi$ to characteristic graph length measures, namely the radius $R$, diameter $D$ and the mean shortest path $\langle s \rangle$. The correlation length is defined as the decay rate of the the two-point correlation function
$G(l)=\langle S_i S_j \rangle - \langle S_i \rangle \langle S_j \rangle$ which measures the correlation of states as a function of the Manhattan distance $l$ between the nodes. Note that $\xi$ characterizes the competition between topology and dynamics on the state of distant nodes, while $R$, $D$ and $\langle s \rangle$ are purely topological measures.

We also address the issue of how connecting two networks of the same size with different rewiring probabilities affects the ordering process. To do so, we compare the results obtained for the two-layer multiplex network ($N$ bonds connecting nodes of two layers in one-to-one fashion) with the results for the case where the same number of inter-layer connections is distributed between randomly chosen pairs of nodes.

The main quantity of interest is the fraction of configurations that have not reached the ground state ("active configurations") $f_a$ after a given simulation time $T$ as a function of $p$. The absolute value of net magnetization $|m|$ is an order parameter for individual systems: $|m|=1$ corresponds to the ground case, whereas $|m|=0$ corresponds to the case in which there is an equal number of spins in both states. Thus, we measure the dependence of the final value of the magnetization $|m_f|$ in disordered configurations on $p$. However, $|m_f|$ contains no information about clustering in the network.

We simulate the dynamics on networks consisting of $50 \times 50$, $80 \times 80$ and $150\times150$ nodes for fixed values of $N$, $\langle k \rangle$, $p$ and $m_0$. The total number of trials in each particular case is set to 1000.
In summary, our numerical algorithm consists of the following steps:

I \textit{Regular network initialization}. Construct lattices with $k=4$ or $k=8$ as in Fig. \ref{fig:lattice_neighbors}.

II \textit{Rewiring}. Following the method described in \cite{herrero2009}, our rewiring process ensures that there are no self-loops or multiple links between pairs of nodes, and that the minimal connectivity degree is 2. Bonds are sequentially selected at random and rewired with probability $p$ until a desired fraction $p$ of the total number of bonds is rewired.

III  \textit{Spin state initialization}. The initial state is set by randomly putting each of the $N$ spins into one of the possible states. If the initial magnetization is $m_0$, the state of each spin is set to $+1$ with the probability $p_{spin}=\frac{1+m_0}{2}$ and to $-1$ with probability $1-p_{spin}$ .

IV \textit{Glauber dynamics}. The evolution of the system is governed by the original or the modified (non-stochastic) Glauber dynamical rule, proceeding either until it reaches the ground state or until it fails to do so after a predetermined number of steps. We choose this value to be $T = 5000N$ ($5000$ attempted spin flips per node).

In what follows, we first analyze the case of a monolayer Watts-Strogatz network, and then consider the ordering process in paradigmatic two-layer networks with two types of inter-layer connections.

\textit{Monolayer Networks -- }Figure~\ref{fig:fa_vs_p}(a) shows how the fraction of active configurations $f_a$ depends on $p$ for Watts-Strogatz
networks with local Glauber dynamics following a zero-temperature quench ($m_0=0$). The nonlinear dependence of $f_a$ on $p$ is observed regardless of $\langle k \rangle$, but turns out to be qualitatively different for the cases $\langle k \rangle=4$ and $\langle k \rangle=8$. When $\langle k \rangle=8$, with increasing randomness ($p \gtrapprox 0.5$), the dynamics leads to almost complete ordering. Nevertheless, when $\langle k \rangle=4$, a finite fraction of configurations fails to reach the ground state in the thermodynamic limit over the whole range of $p$ values. In the small-world regime, however, the ground state is not reached in the thermodynamic limit in either case. The result that ordering cannot be attained in small-worlds when state transitions are governed by Glauber dynamics has been previously demonstrated for rewired rings ($d=1$) and rewired square lattices ($d=2$) with $\langle k \rangle=4$
\cite{herrero2009,boyer2003}.

\begin{figure}
\centering
\includegraphics[scale=0.35]{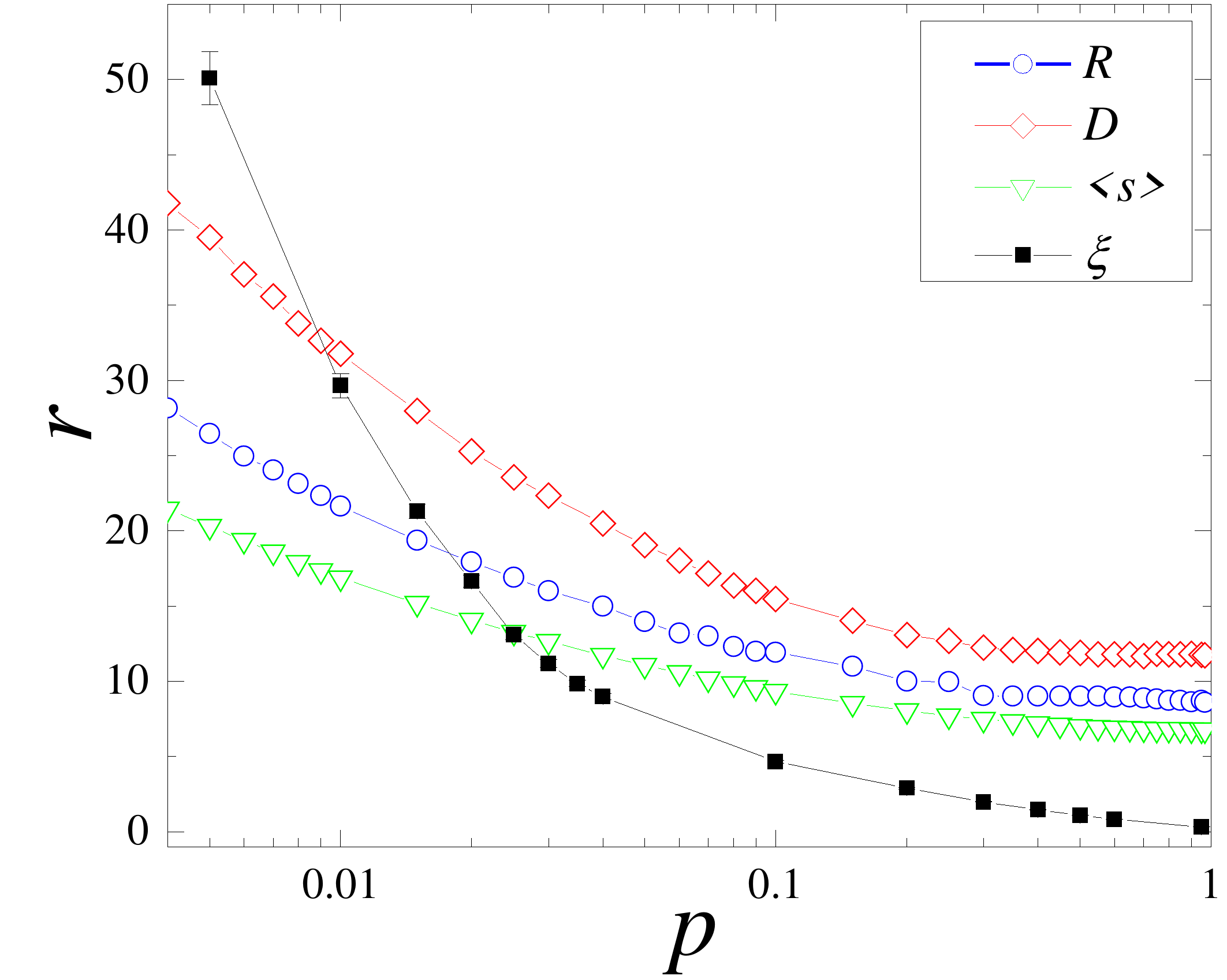}
\caption{(Color online) Correlation length $\xi$ compared to graph distance measures (radius $R$, diameter $D$, and average path length $\langle s \rangle$) as functions of $p$. While $\xi$ reflects the interplay between the dynamics and the network structure, the remaining quantities characterize purely topological features of the network. Crossing of $\xi(p)$ with other curves indicates the transition between the dynamics typical for the regular lattices and that for the rewired networks. Note that all four quantities exhibit a power law dependence
$r \propto p ^{-a}$ in the $p$ region approximately coinciding with the small-world regime. The results refer to
networks with $80\times80$ nodes and $\langle k \rangle=4$.} \label{fig:corr_length}
\vspace{-0.6cm}
\end{figure}

One infers that the local neighborhood majority rule with stochastic spin flips cannot lead to an ordered state on graphs with a perturbed regular topology. While the neighborhood from the regular lattice is mostly conserved in the small-world limit, $R$, $D$ and $\langle s \rangle$ on the other hand monotonically decrease with $p$ due to the presence of shortcuts (see Fig. \ref{fig:corr_length}). Thus, it follows that perturbing the local neighborhood essentially leads to dynamical frustration of the local majority rule. A very small amount of topological disorder is sufficient to induce the critical slowing down of dynamics, causing the disordered states to appear as deformed stripes on the lattice. Further deformation of the stripes leads to multiclustering on the lattice, which is reflected in the crossover effect \cite{barthelemy1999}. We have established that this effect corresponds to the drop of $\xi$ below the topological distances. At the same time, the low value of $\xi$ indicates the absence of long-range ferromagnetic order. The two-point correlation function is found to satisfy an exponential scaling law $G(l) \propto e^{-\frac{l}{\xi}}$ over the whole range of $p$. Furthermore, depending on the $p$ value, both $\xi$ and
$R$, $D$ and $\langle s \rangle$ exhibit different scaling regimes.

\begin{figure}
\centering
\includegraphics[scale=0.35]{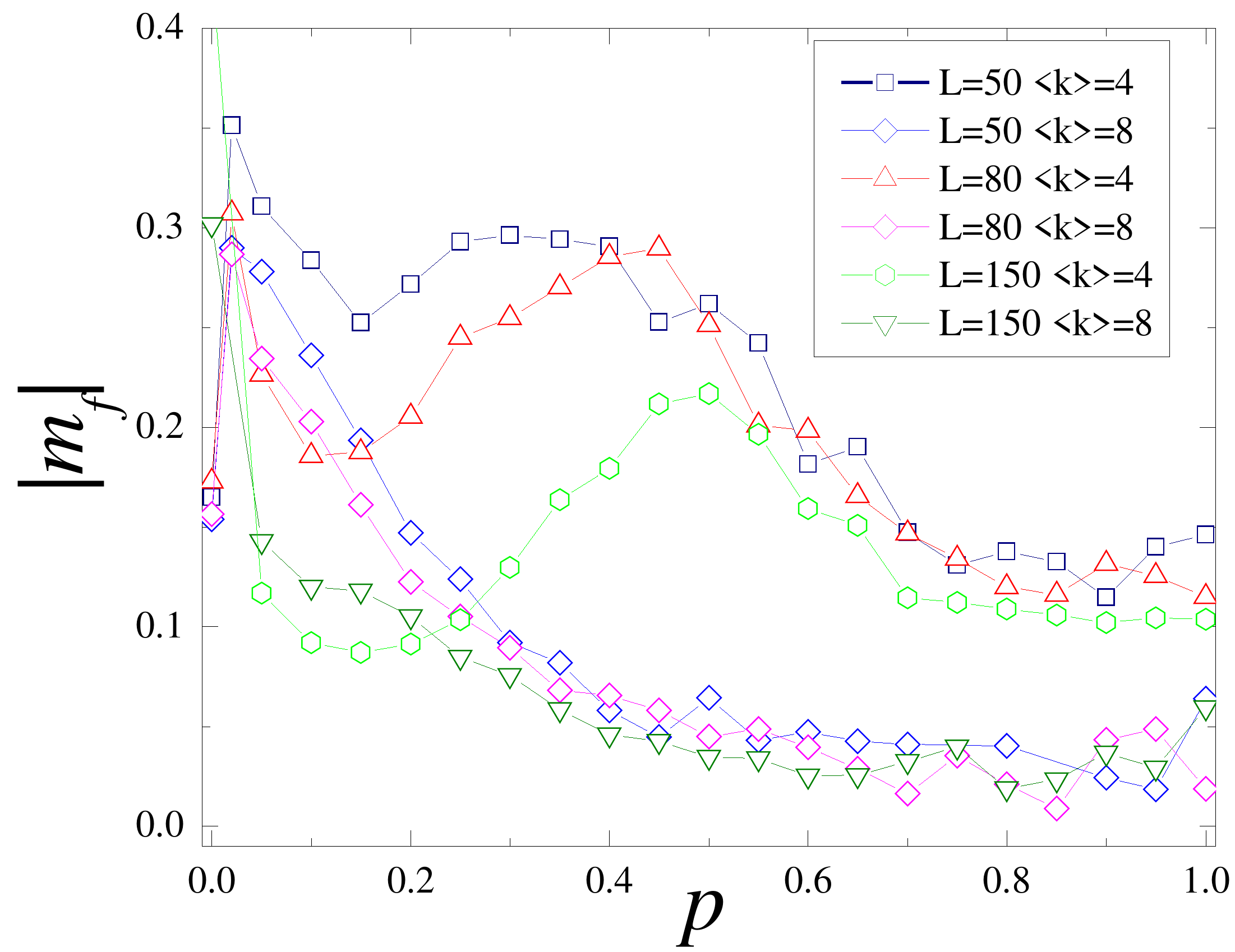}
\caption{(Color online) Final magnetization averaged over the ensemble of disordered configurations $|m_f|$ in dependence of $p$. For $\langle k \rangle=8$, one finds approximately equal numbers of nodes in both states as
$p \rightarrow 0$. For $\langle k \rangle=4$, within the small-world regime, $|m_f|$ is reduced compared to the regular lattice, while for intermediate $p$, the droplet configurations lead to increase of $|m_f|$. The peak of $|m_f|(p)$ gets shifted because the fraction of active runs is higher at a wider range of $p$ values for larger networks, cf.
Fig. \ref{fig:fa_vs_p}(a).} \label{fig:magnet_actconfs}
\vspace{-0.6cm}
\end{figure}

\begin{figure*} [t]
\centering
\includegraphics[scale=0.36]{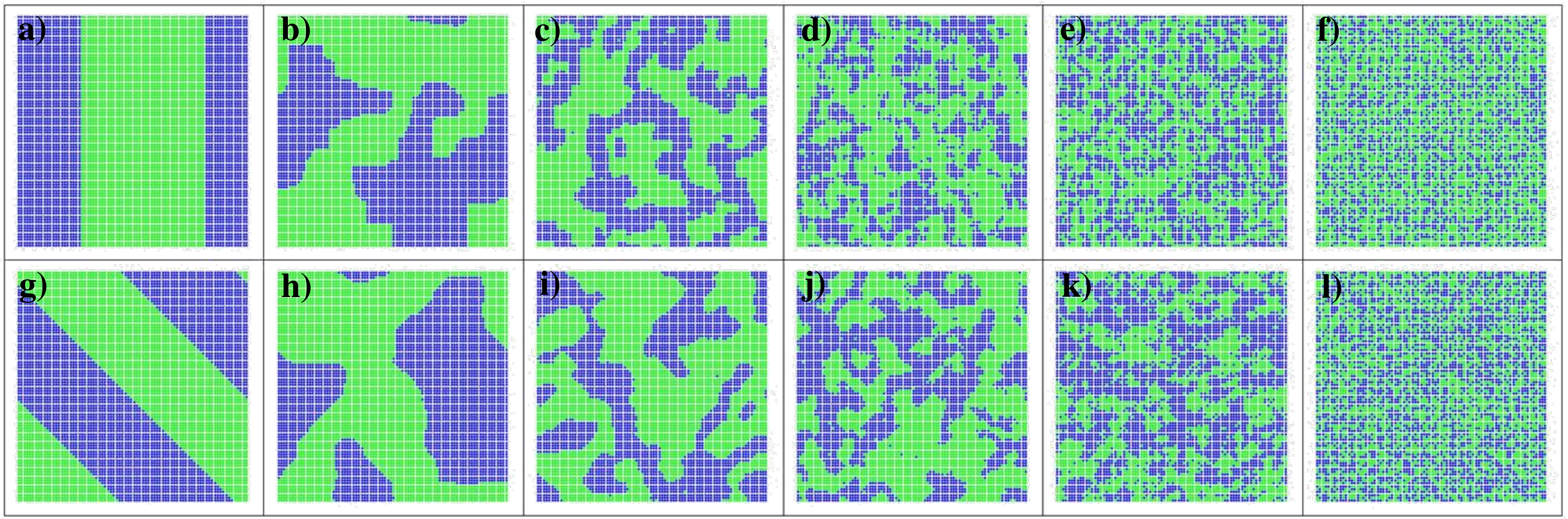}
\caption{(Color online) Snapshots of disordered configurations at $T=T_f$ on the lattice. The top (bottom) row refers to networks with $\langle k \rangle=4$ ($\langle k \rangle=8$). The rewiring probabilities are $p=0$ in (a) and
(g), $p=0.02$ in (b) and (h), $p=0.1$ in (c) and (i), $p=0.3$ in (d) and (j), $p=0.5$ in (e) and (k), as well as $p=1$ in (f) and (l). The stripe structure is gradually lost with increasing $p$, giving way to the multiclustered states with respect to the original lattice. The number of domains increases with $p$ as the network topology substantially departs from the lattice one. In terms of the network structure, each of the disordered configurations consists of two connected components. All the results are obtained for networks with $80 \times 80$ nodes.} \label{fig:final_snapshots}
\end{figure*}

In particular, in the small-world regime, $R, D, \langle s \rangle$ and $\xi$ exhibit a power law dependence on $p$, $r \propto p^{-a}$ with $r\in\{R,D,\langle s \rangle,\xi\}$ and $a\in\{a_R,a_D,a_{\langle s \rangle},a_{\xi}\}$. For $80 \times 80$ networks, the following values for the exponent $a$ are found: $a_{R} = -0.259 \pm 0.004$, $a_{D} = -0.296 \pm 0.005$, $a_{\langle s \rangle} = -0.25 \pm 0.003$ and $a_{\xi} = -0.77 \pm 0.01$. For larger values of $p$, the topological measures do not change significantly with increasing $p$ indicating that topological effects remain the same after $\approx 0.5$. Nevertheless, $\xi$ decays to zero as $p \rightarrow 1$, which implies that the dynamics is sensitive to rewiring over the whole range of $p$, as corroborated by the growing number of "clusters" of decreased sizes in disordered configurations for large $p$, see Fig. \ref{fig:final_snapshots}.

Interestingly, a deeper understanding of the difference in ordering efficiency in terms of $p$ may be gained by considering $f_a$ for configurations governed by the modified Glauber rule. Evidently, the difficulty in attaining order subsides when stochasticity is eliminated from the dynamics in the small-world limit regardless of
$\langle k \rangle$, see Fig. \ref{fig:fa_vs_p}(b). In other words, the ground state is reached with probability one
if energy-neutral state transitions are not allowed. This always holds for $\langle k \rangle=8$, and also for networks with $\langle k \rangle=4$ in the limits $p \rightarrow 0$ and $p \rightarrow 1$. For intermediate $p$, ordering remains suppressed to a certain degree.

\begin{figure}
\includegraphics[width= \linewidth]{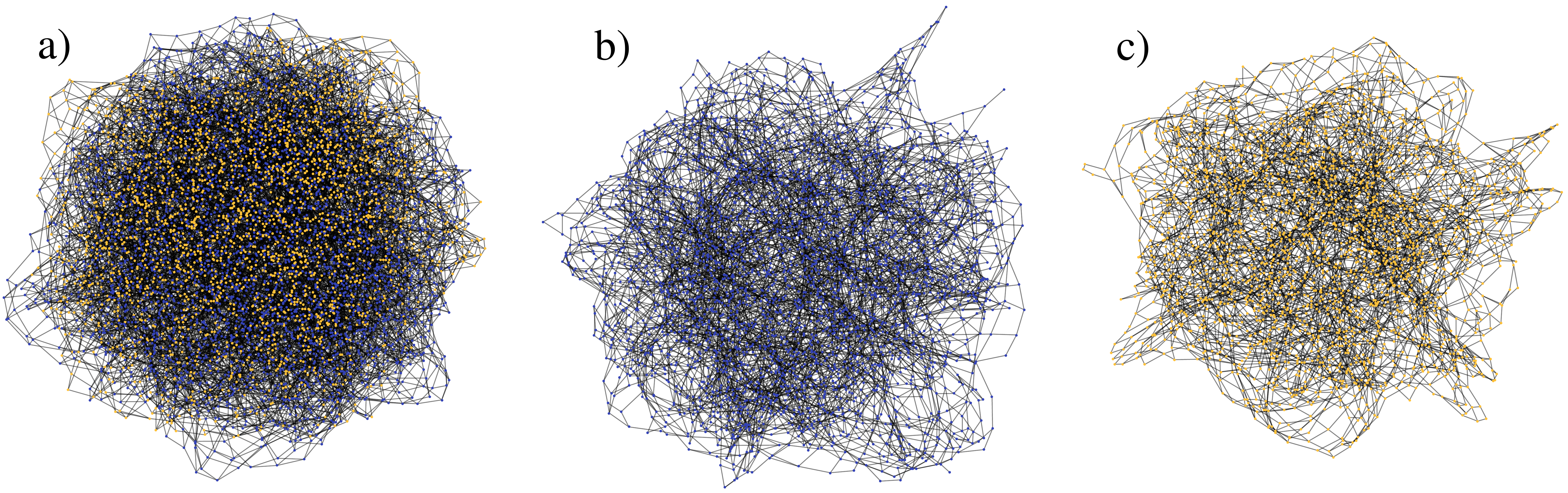}
\caption{(Color online) Example of a disordered configuration obtained for the network of size $80 \times 80$, with $\langle k \rangle=4$ neighbors on average and $p=0.1$ rewired links. (a) refers to the full configuration, whereas (b) and (c) show the larger component (3637 nodes) and the smaller component (2763 nodes), respectively. The final magnetization is $|m_f| \approx 0.14$. The nodes are separated into two domains of similar size, forming a multi-domain state on the lattice, cf. Fig. \ref{fig:final_snapshots}(c).}\label{fig:3d_partitioning}
\vspace{-0.6cm}
\end{figure}

The next objective is to demonstrate that varying initial magnetization $m_0$ allows one to interpolate between the influences of dynamics and topology. Fig. \ref{fig:fa_vs_p}(c) shows $f_a$ as a function of $p$ for $m_0 \neq 0$ under the standard Glauber rule. While initial bias towards local clustering promotes complete ordering for regular networks, the dynamical outcome is different for rewired networks. In case $\langle k \rangle=8$, small values of $m_0 \neq 0$ significantly increase ordering, whereby the position of the peak of $f_a(p)$ coincides with the peak value of $f_a(p)$ at $m_0=0$. Perturbing the quenched initial state on graphs in the small-world regime increases the prevalence of the ground state. Nevertheless, the peaks of $f_a(p)$ curves for $\langle k \rangle=4$ networks in Fig. \ref{fig:fa_vs_p}(c) shift toward the peak value from Fig. \ref{fig:fa_vs_p}(b) as $m_0$ is increased. A fraction of configurations still fails to reach the ground state for some values of $p$, even for high values of $m_0$. The shift demonstrates that as the number of stochastic state transitions decreases due to the initial bias in clustering, the dynamical frustration is reduced. Nonetheless, the topological obstructions in networks with low $\langle k \rangle$ can suppress ordering even for high values of $m_0$.

Further insight on this issue can be gained by observing how $|m_f|$ averaged over active configurations depends on $p$, see Fig. \ref{fig:magnet_actconfs}. The initial increase in magnetization corresponds to the divergence of relaxation time in the limit $p \rightarrow 0$, i.e. the crossover from large-world to small-world behavior. Expectedly, there is a qualitative difference in the $|m_f|(p)$ profile for different $\langle k \rangle$ under increasing $p$. The curves for $\langle k \rangle=8$ monotonically decrease, indicating that the small number of configurations that does survive converges to a state consisting of a similar number of opposite spins in the limit $p \rightarrow 1$. In contrast, the initial decrease in $|m_f|$ for networks with $\langle k \rangle=4$ is followed by the peak at intermediate values of $p$, associated to the presence of droplet configurations with $|m_f| \rightarrow 1$.

\begin{figure*}[t]
\centering
\includegraphics[scale=0.6]{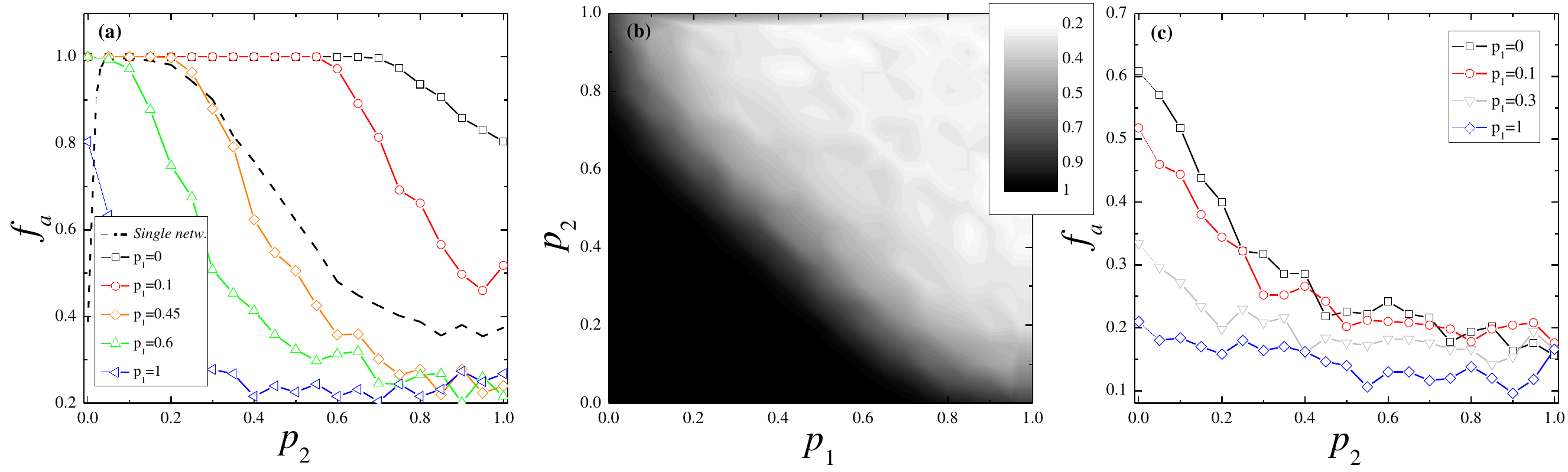}
\caption{(Color online) (a) Ordering in two-layer multiplex networks: $f_a$ for the layer with rewiring probability $p_1$ (see the legend) in terms of the rewiring probability $p_2$ of the other layer. For comparison, the results for the single (monolayer) network are indicated by the dashed line. (b) $f_a$ of a single layer of a multiplex network as a function of $p_1$ and $p_2$. The states of two layers are strongly correlated, but the ordering is completely inhibited in the small-world regime. (c) $f_a$ of a layer of the network with random inter-layer connectivity. Ordering is significantly improved for all values of $p$. All the results are obtained for networks with $80 \times 80$ nodes and $\langle k \rangle=4$.} \label{fig:multiplex_and_random}
\vspace{-0.4cm}
\end{figure*}

Moreover, we have verified that the disordered configurations in the small-world regime consist of two intertwined topological spin domains of almost similar size with stochastically fluctuating interfaces. An example of such a two-component state for $p=0.1$ is provided in Fig. \ref{fig:3d_partitioning}, whereby the corresponding lattice domain configuration is shown in Fig. \ref{fig:final_snapshots}(c). Blinkers that arise as a result of the long-range connections can be present along with stochastic flipping of interfaces on the lattice. Increasing $p$ corresponds to the formation of domains with decreasing size with respect to the lattice. Several examples of configurations with two topological components for different $p$ are shown in Fig. \ref{fig:final_snapshots}. The number of these domains counted on the lattice grows exponentially with $p$ (not shown). In the random network limit, as the fraction of links belonging to the original lattice $1-p$ decreases, clusters become indistinguishable when observed on the lattice. Topologically, the two-domain configuration is reminiscent of the disordered configurations of the voter model on small-world networks \cite{castellano2003,castellano2005}. Once the dynamics cannot cause further decrease in energy, the interface length reaches a constant value, as interface diffusion is no longer possible. In this scenario, while a fraction of nodes with even connectivity degrees continues to flip indefinitely with no energy cost, the states of the odd-degree nodes become stationary.

Nevertheless, the disordered configurations associated to the increase of $f_a$ in figure \ref{fig:fa_vs_p}(b) for $\langle k \rangle=4$ are frozen at very high values of $|m_f|$, viz. $|m_f| \rightarrow 1$ in the thermodynamic limit, and correspond to \textit{absorbing} states of the network. These configurations result from low connectivity and consist of a tiny fraction of spins isolated in small domains surrounded by the "sea" of nodes of the opposite orientation. In this scenario, nodes with small $k$ form stable droplets of opposing magnetization which cannot be dynamically influenced by the nodes from the rest of the network, preventing the system from reaching the full order. These droplets may appear on the remnants of the regular lattice, such that their interior consists of nodes connected by links from the regular lattice ($k=4$), whereas their boundary is mainly comprised of nodes with one removed link ($k=3$), thereby trapping the "interior" in the same state. Even-degree nodes that appear on the boundaries have more links with nodes within the droplet than with other neighbors, such that their state cannot be changed either. With further rewiring of the lattice, stable droplets may still form as even smaller groups of interconnected nodes with small degrees ($k=2$ or $k=3$), likewise disconnected from the rest of the network. The larger the network, the more likely becomes such a scenario. Also, for larger network sizes, a larger number of droplets may be present, which is the reason for why a larger fraction of configurations fails to reach order. This peculiar frustration on the remnants of a regular lattice also accounts for the incomplete ordering of systems governed by the standard Glauber dynamics in rewired networks with $\langle k \rangle=4$, and explains for the difference in the behavior in the limit $p \rightarrow 1$. Final configurations in networks with $\langle k \rangle=4$ can consist of two large components and a few isolated droplets for $p$ above the small-world regime, similar to final configurations obtained for $m_0 \neq 0$.

\textit{Two-layer networks -- }We now address the ordering process in multilayer networks, focussing on the paradigmatic example of two coupled $\langle k \rangle=4$ networks with different rewiring probabilities $p_1$ and $p_2$. By our algorithm, the individual layers are rewired consecutively, after which $N$ links are introduced between them, either at random or with the one-to-one correspondence between the layers' nodes. The simulation is terminated after $2T$ steps if order is not reached. Note that introducing new links effectively generates a large network with $\langle k \rangle=5$ and $5N$ bonds.

Our findings indicate that both cases lead to highly correlated states of layers, which are simultaneously ordered/disordered and have $m_{f 1} \approx m_{f 2}$. For this reason, $f_a$ of a single layer presents an appropriate quantity to characterize the ordering process. We find that the dependence of $f_a$ on rewiring probability changes qualitatively depending on the nature of the inter-layer bonds, cf. Fig. \ref{fig:multiplex_and_random}(a) and Fig. \ref{fig:multiplex_and_random}(c). The multiplex configuration turns out to suppress ordering of both networks in the small-world regime, as indicated in Fig. \ref{fig:multiplex_and_random}(a) and Fig. \ref{fig:multiplex_and_random}(b). However, Fig. \ref{fig:multiplex_and_random}(a) shows that ordering efficiency can be increased if at least one of the networks is "sufficiently random", with a smooth transition taking place at $0.35 \lesssim  p \lesssim  0.45$. Interestingly, the other scenario, which involves placing the same number of bonds between randomly chosen pairs of nodes from both networks, promotes both ordering and correlation between the layer states. This is corroborated by Fig. \ref{fig:multiplex_and_random}, suggesting that regardless of $p_1$ and $p_2$, ordering in this case is significantly improved compared to that on a single network and the multiplex network.

The curves obtained for multiplex networks resemble the ones obtained for the single network even for $p$ as large as $0.6$, while those obtained for random inter-layer connections are monotonically decreasing as $p_2$ is increased over the whole range of $p$. Even though in both cases networks become correlated in terms of $m_f$ and ordering, multiplexing seems to preserve the type of dynamics obtained on small-world structures of one network, while introducing random bonds between the layers destroys the small-worldliness effect.

\textit{Conclusions -- }We have analyzed ordering efficiency of the Glauber model of Ising spin kinetics on the Watts-Strogatz networks obtained by rewiring from the two-dimensional square lattices with coordination numbers $\langle k \rangle=4$ and $\langle k \rangle=8$. We have extended the previous results concerning the failure of such systems to reach the ground state in the small-world regime $0<p \ll 1$, gaining insight into the associated disordered configurations.
The fraction of active configurations exhibits a nonlinear dependence on the rewiring probability. It is interesting that the similar type of dependence has been observed in relation to synchronization process on small-world networks \cite{grabow2010}.
It is found that the Glauber dynamics on small-world networks becomes stuck in metastable stationary active configurations, which consist of two intertwined domains of opposite spins, whereby the fraction of nodes on the interfaces flips indefinitely. This effect is manifested as clustering patterns in the lattice representation. The size of domains on the lattice becomes smaller as $p$ is increased. We have demonstrated that the limiting value of $p$ at which the number of lattice and topological domains is equal (to two) corresponds to the value where the correlation length $\xi$ becomes smaller than the average path length in the network.

Our analysis shows that the active configurations in the small-world regime emerge when the perturbed regular topology constrains the number of possible energy lowering processes, while the stochastic energy-neutral spin-flipping processes contribute to dynamical frustration and trap the system in a set of metastable states with the same energy. While the ground state is not accessible because energy lowering processes are not possible, the energy-neutral processes allow for the transitions between the states of the same energy. This is similar to what has been reported for Glauber dynamics on $3D$ regular lattices \cite{spirin2001-2}, Glauber dynamics on random graphs \cite{castellano2005}, and the voter model on small-world networks \cite{castellano2003}.

We have further demonstrated that there exists a finite probability of finding another type of disordered configuration in networks with low connectivity for intermediate values of $p$. These are frozen, almost completely ordered states with a few isolated droplets of opposing magnetization. For $\langle k \rangle=8$, such configurations become unlikely due to the high average connectivity degree in the network, giving way to fully ordered states if $p$ is sufficiently increased ($p>0.5$). In networks with $\langle k \rangle=4$, a certain fraction of configurations exists as a combination of these states, especially if an initial bias towards clustering ($m_0 \neq 0$) is introduced.

We have also examined the features of the ordering process in paradigmatic two-layer networks. It has been found that the structure of inter-layer connections strongly affects the ordering process. In particular, multiplexing decreases ordering efficiency in the small-world regime $0<p \ll 1$, but improves it if the rewiring probability in both layers is sufficiently high. Nevertheless, random connectivity between the layers always promotes ordering, regardless of layer topology. In all the considered scenarios, the layers typically end up in highly correlated states.

We believe that the future research may be directed towards extending our findings on the dynamics of interacting rewired networks. In particular, it could be interesting to modify inter-layer coupling strengths, vary the number of connections between the layers or consider hierarchical networks and networks with a large number of layers.

\acknowledgments
This research was supported by the Ministry of Education, Science and Technological Development of
Republic of Serbia (Grant 171017) and by the Slovenian Research Agency (Grants J1-7009 and P5-0027).

\end{document}